# On the ionization equilibrium of hot hydrogen plasma and thermodynamic-consistency of formulating finite internal partition functions

Mofreh R. Zaghloul

Department of Physics, College of Sciences, United Arab Emirates University,

P.O.Box 17551, Al-Ain, UAE.

**ABSTRACT**

The problem of formulating a thermodynamically-consistent finite internal partition function (IPF) in nonideal hydrogen plasma systems is investigated and analyzed within the chemical picture revealing inaccuracies and inconsistencies buried in widely used formulations in the literature. The analysis carried out here, though performed for the simplest case of pure nonideal hydrogen plasma, it shows all specific features of the problem and is extendable to the general case of a complex multi-component plasma mixture. A criterion for the separability of the configurational component of the free energy is presented and an accurate and consistent formulation of the problem is introduced. The presented criterion and the introduced consistent formulation of the problem clear ambiguities in other formulations in the literature and provide a better understanding of the problem. An illustrative example is worked out showing simplicity and effectiveness of the proposed consistent formulation and the importance of terms, essential for thermodynamic consistency, which are commonly neglected by other authors in the literature.

# I- INTRODUCTION

In the chemical picture of plasmas, both of composite particles, nuclei and free electrons are treated as elementary members of the thermodynamic ensemble. At a fixed volume and constant temperature, a physical system of particle numbers *{N}* can be statistically described by the canonical partition function or by its associated thermodynamic potential, the Helmholtz free energy function *F*. The condition of thermodynamic equilibrium requires that the free energy must be a minimum and the equilibrium state is determined, therefore, by minimizing the free energy with respect to the occupation numbers *{N}*, subjected to the stoichiometric constraints of electroneutrality and conservation of nuclei. The free energy minimization technique is known to assure thermodynamic consistency among the particle occupation numbers and thermodynamic properties derived from the derivatives of the same free energy function as demonstrated in Ref. [1]. In addition, the technique is shown to be versatile and efficient in studying interacting plasma systems and might probably be the most widespread-technique used in the literature for the calculation of the equation-of-state, thermodynamic properties and radiative characteristics of laboratory and space plasmas (see for example Refs. [1-5]). High density effects and interactions among different species in the system cause a departure from the ideal behavior which is usually taken into account in terms of a modified free energy function. The factorizability of the translational, configurational and internal components of the total many-body partition function and its equivalent separability of the corresponding components of the associated Helmholtz free energy function have been assumed by many authors in the literature (see for example Hummer and Mihalas (HM) in Ref. [2] and Potekhin in Ref. [5]) such that

$$F\ (\{N_k\}, V, T) = F_{trans} + F_{conf} + F_{int} \qquad (1)$$

where *V* is the volume of the system, *T* is the absolute temperature and *{N}* refers to the occupation numbers of different plasma species. The terms $F_{trans}$, $F_{conf}$ and $F_{int}$ in Eq. (1) refer to the total translational, configurational and internal components of the free energy, respectively.

Various approaches used for the minimization of the free energy function (see for example Refs. [1-7]) would require, however, a thermodynamically-consistent scheme for establishing finite internal partition functions IPFs for composite particles. The use of the temperature-dependent Planck-Larkin partition function (PLPF), for nonideal plasma systems has been critically criticized by Rouse in Ref. [8] who argued that the use of PLPF for nonideal plasma systems suffers major physical problems. In a response to Rouse's criticism, Ebeling et al in Ref. [9] recommended, for quantum statistical Coulomb system with bound states, the use of the discrete energy states of the Bethe-Salpeter equation (BSE) in the PLPF in for the later to become temperature-and density-dependent. Following to Ebeling et al response to Rouse's criticism, Rogers in Ref. [10] and Däppen et al (see, for example Refs [11, 12]) stressed that despite its name the PLPF is not a true partition function but merely an auxiliary term in a virial coefficient. Accordingly, the present analysis and discussion will be concerned with formulations of the partition functions which are both temperature and density-dependent. It has to be recalled, however, that the approach used for establishing finite IPFs should properly imitate the mechanism by which bound states are swept to the "continuum" causing the converging and finiteness of the IPFs.

**II- THE OCCUPATIONAL PROBABILITY FORMALISM**

The central problem of thermodynamic equilibrium at high density, within the chemical model, lies in the consistent determination of the density and temperature dependence of the scheme used to formulate finite internal partition functions of the ensemble of bound states. The abrupt cutoff or abrupt truncation of the internal partition function at some maximum state-dependent number of bound states (or energy levels) (see for example [13-17]) in order to account for finite-density effects, though can stand on some physical considerations, is known to suddenly switch a state from being "bound" to "free". This abrupt disappearance of the bound state into the continuum turns the IPF into a nonanalytical function of its variables and produces undesirable discontinuities and singularities in the free energy and its derivatives in the expressions of the thermodynamic functions. The situation gets even worse with *ad hoc* cutoff schemes in which the partition function is truncated just after the ground level (see, for example Ref. [18]) or when the atomic states are grouped into a few virtual levels and the calculation of the partition function considering these virtual levels to follow Boltzmann distribution as recently suggested in Ref. [19].

One of the widely used ways to treat these inconsistencies and to establish a finite internal partition function IPF in the nonideal plasma environment while avoiding these discontinuities consists in using the occupation probabilities. They were used by many authors, starting with Fermi in 1924 (Ref. [20]), but have become most popular in the calculation of equation of state, thermodynamic properties and radiative characteristics of laboratory and astroplasmas after the papers by Hummer, Mihalas and Däppen (see for example, Refs. [2,3]). The approach of the occupation probability formalism is extensively adopted and used in the literature (see for example Refs. [5,21-25]).

In the occupation probability formalism, weights are assigned to all bound states of all composite species in the system. The internal partition function of the ion $\zeta$ can therefore be written as

$$Q_{int,\zeta}(\{N_{\zeta,k}\},V,T) = \sum_k w_{\zeta,k}(\{N_{\zeta,k}\},V,T) \exp\left(\frac{\chi_{\zeta,k}}{K_B T}\right)$$
$$= \sum_k g_{\zeta,k}\, \omega_{\zeta,k}(\{N_{\zeta,k}\},V,T) \exp\left(\frac{\chi_{\zeta,k}}{K_B T}\right) \quad (2)$$

where $V$ is the volume of the system, $K_B$ is the Boltzmann constant, $N_{\zeta,k}$, $g_{\zeta,k}$ and $\chi_{\zeta,k}$ represent the occupation number, the statistical weight and the unperturbed binding energy of the $k$th quantum state of the ion $\zeta$, respectively. The state-dependent occupational probability $w_{\zeta,k}(\{N_{\zeta,k}\},V,T) = g_{\zeta,k}\, \omega_{\zeta,k}(\{N_{\zeta,k}\},V,T)$ of the $k$th quantum state is presumed to decrease continuously and monotonically as the strength of the relevant interactions increases in order to produce a physically reasonable continuous transition between bound and free states. A natural and smooth truncation of the internal partition function would also require that the occupational probability $w_{\zeta,k}(\{N_{\zeta,k}\},V,T)$ or equivalently the weight $\omega_{\zeta,k}(\{N_{\zeta,k}\},V,T)$ to fall strongly to zero as the binding energy of a level below the unperturbed continuum goes to zero.

It has to be noted however that, as shown by Fermi in Ref. [18] and stressed by HM in Ref. [2] and Potekhin in Ref. [5], the introduction of an occupational probability $w_{\zeta,k}(\{N_{\zeta,k}\},V,T)$ necessitates a modification of the free energy such that $w_{\zeta,k}(\{N_{\zeta,k}\},V,T)$ becomes consistent with and can be derived from the adopted form of the configurational component of the free energy, $F_{conf}$. A common factor among HM and Potekhin's formulations is the assumed separability of the configurational component of the free energy and the assumed possibility of deriving corresponding occupational probabilities, $w_{\zeta,k}(\{N_{\zeta,k}\},V,T)$, that depend on the occupation numbers of the individual excited states, for any form of the configurational component of the free energy; a hypothesis that will be strongly questioned in the present study.

Considering the simplest case of a single species perfect gas of *neutral* particles (ionization processes and charged particles are not taken into account) HM in Ref. [2] assumed a separable term $f = F_{conf}$ to represent the configurational component of the free energy that is supposed to depend explicitly on the occupation numbers $\{N_k\}$ of the individual excited states. Implementing equilibrium between excitation/de-excitation processes they derived for the occupational probability the form

$$w_k = g_k \, \omega_k \equiv exp\left(-\frac{\partial f/\partial N_k}{K_B T}\right) \quad (3)$$

However, upon using the form (3) for $w_k$ and substituting back into the free energy function they recovered a residual term

$$\left[f - \sum_k N_k (\partial f / \partial N_k)\right] \quad (4)$$

which contradicts the assumed separability of the configurational component of the free energy in the case of using an occupational probability in the form (3). Reusing the new term in Eq. (4) will, in general and except for the special case in which $f$ is linear in $\{N_k\}$, lead to a different form for $w_k$ that depends on higher order derivatives of $f$ with respect to $\{N_k\}$ and so on. Hummer and Mihalas had to assume linear dependence or astute linearization of the interaction term on $\{N_k\}$ and they made every possible approximation in an attempt to satisfy such a linear dependence even with the inclusion of ionization and the involvement of the long range Coulomb forces among the charged particles. Such a difficult-to-justify linearity of $f$ with respect to $\{N_k\}$ was never fully satisfied even for the simplest way to treat interactions among neutral particles, using the hard sphere model, in which a definite radius is assigned to each particle species. As stated in their paper, the HM linearization leads to an unavoidable double counting with an *exponent* of the occupation probability smaller by *a factor of 2*. It has also to be noted that the use of $w_k$ as given by HM in conjunction with unperturbed energy levels in the calculation of HM partition functions does not formally ensure convergence or even reasonable convergence of the IPF.

In a following work, Potekhin in Ref. [5] made an attempt to modify or extend the occupational probability formalism for solving the ionization equilibrium problem. Considering the case of hydrogen plasma including charged particles (i.e; taking ionization processes into consideration) and assuming the separability of the configurational component of free energy, Potekhin derived a form for the occupational probability that takes the form

$$w_k \equiv g_k \; exp\left(-\frac{1}{K_B T}\left(\frac{\partial F_{conf}}{\partial N_k} - \frac{\partial F_{conf}}{\partial N_p} - \frac{\partial F_{conf}}{\partial N_e}\right)\right) \tag{5}$$

where $N_p$, $N_e$, and $N_k$ are the occupation numbers of protons, electrons and hydrogen atoms at quantum state *k*, respectively. Eq. (5), which reduces to Eq. (4) in the absence of charged particles, was introduced as a form of the occupational probability that is thought to be consistent with, and can be derived from, any proposed form of separable configurational component of the free energy function. A restriction of linear-dependence of $F_{conf}$ on $\{N_k\}$, similar to what has been affirmed by HM, has never been stated or declared by Potekhin and the formulation (5) was confusingly introduced as a "*general*" or restriction-free formulation.

In the present paper, we explain that the use of a separable configurational component of the free energy, leading to a form of $w_k$ similar to those derived by HM or Potekhin, is *strictly conditional* for very limited situations of little or no practical importance and usefulness. Besides, we introduce the criterion or the required condition for such separability in case of using a state-dependent partition function for occupational probability formalisms similar to those by HM and Potekhin. In addition, an accurate, formal solution of the problem is introduced in terms of the solution of the inverse problem in which an occupational probability (or a scheme to establish finite IPFs) can be assumed in advance enabling the solution for the occupation numbers of particle species; the corresponding set of modified and consistent thermodynamic functions are derived and presented. Needless to say that the adopted or developed model for the occupational probability (or the scheme used to establish finite IPFs) is supposed to rely on the physics of the particle interactions ensuring the convergence or finiteness of the internal partition function.

Values of the corrected or modified free energy of the system, if needed, can be found using the computed occupation numbers either analytically or numerically depending on the adopted form of the occupational probability (or the scheme used to establish finite IPF).

## III- INCONSISTENCIES IN POTEKHIN'S FREE ENERGY FORMULATION

Our interest in this section is to point out and to explain the inaccuracy/inconsistency in Potekhin's formulation. To identify the inaccuracy/inconsistency buried in Potekhin's formulation, we consider a system of pure hydrogen in a temperature range where a considerable fraction of atoms can exist in excited states and the convergence of the IPF becomes crucial. Adopting the same assumption of fully dissociated molecules, as in Potekhin's work, the plasma system is essentially composed of electrons, protons and hydrogen atoms. The free energy models of these components are discussed below.

### A- Electron's free energy and degeneracy

At high densities, the degeneracy of free electrons has to be taken into account, while protons and neutral hydrogen atoms remain classical because of their relatively heavy masses. The free energy of electrons can therefore be written as

$$F_e^{dgc} = N_e K_B T \left[ ln\left(\frac{N_e \lambda_e^3}{2V}\right) - 1 \right] + \Delta F_e^{dgc} \qquad (6)$$

where $N_e$ is the occupation number of electrons and $\lambda_e = h/\sqrt{2\pi m_e K_B T}$ is the electron's thermal de Broglie wavelength. The term $\Delta F_e^{dgc}$ refers to the quantum or degeneracy correction to the classical free energy which can be formally written as

$$\Delta F_e^{dgc} = \left(F_e^{dgc} - F_e^{clc}\right) \qquad (7)$$

where the classical free energy for electrons, $F_e^{clc}$, is given by the first term in the r.h.s. of Eq. (6). Potekhin used for the term $\Delta F_e^{dgc}$, the approximate expression

$$\Delta F_{e,PK}^{dgc} = N_e K_B T \left(\frac{N_e \lambda_e^3}{V \, 2^{7/2}}\right) \qquad (8)$$

However, the rigorous expression for $\Delta F_e^{dgc}$ can be expressed as in Ref. [2,3,26,27]

$$\begin{aligned} \Delta F_e^{dgc} &= -P_e V + N_e \mu_e - F_e^{clc} \\ &= -\left(2 K_B T V / \lambda_e^3\right) I_{3/2}(\mu_e/K_B T) + N_e \mu_e - F_e^{clc} \end{aligned} \qquad (8')$$

In Eq. (8') $P_e$ and $\mu_e$ represent the pressure and chemical potential of the ideal Fermi electron gas, respectively, and $I_\nu$ is the complete Fermi-Dirac integral of the order $\nu$,

$$I_\nu(x) = \frac{1}{\Gamma(\nu+1)} \int_0^\infty \frac{y^\nu}{exp(y-x)+1} \, dy \qquad (9)$$

where $\Gamma$ is the gamma function. The electron chemical potential $\mu_e$ is related to the number of free electrons in the system by

$$N_e = \left(2V/\lambda_e^3\right) I_{1/2}(\mu_e/K_B T) \qquad (10)$$

Now, and for future use, one can express the partial derivative of the correction $\Delta F_e^{dgc}$ with respect to electrons' occupation number from Potekhin's approximate expression and from the above rigorous expression (Eq. (8')) as

$$\frac{\partial \Delta F_{e,PK}^{dgc}}{\partial N_e} = N_e K_B T \frac{\lambda_e^3}{2^{5/2} V} \tag{11}$$

and

$$\frac{\partial \Delta F_e^{dgc}}{\partial N_e} = K_B T \left[ I_{1/2}^{-1}\left(\frac{N_e \lambda_e^3}{2V}\right) - \ln\left(\frac{N_e \lambda_e^3}{2V}\right) \right] \tag{11'}$$

where $I_\nu^{-1}$ is the inverse Fermi-Dirac integral of the order $\nu$. Therefore the partial derivative of the electron's free energy with respect to $N_e$ can be written as

$$\begin{aligned}\frac{\partial F_e}{\partial N_e} &= \frac{\partial F_e^{clc}}{\partial N_e} + \frac{\partial \Delta F_e^{dgc}}{\partial N_e} \\ &= K_B T \ln\left(\frac{N_e \lambda_e^3}{2V}\right) + \frac{\partial \Delta F_e^{dgc}}{\partial N_e}\end{aligned} \tag{12}$$

**B- Protons' free energy**

Since protons have no bound electrons, the internal partition function of the proton is unity and the total free energy of the protons is effectively their translational free energy where

$$F_p = F_{p,trans} = N_p K_B T \left[ \ln\left(\frac{N_p \lambda_p^3}{V}\right) - 1 \right] \tag{13}$$

where $\lambda_p = h/\sqrt{2\pi m_p K_B T}$ is the proton's thermal de Broglie wavelength. Similarly, one can use Eq. (13) to find the partial derivative of the protons' free energy with respect to the protons' occupation number where

$$\frac{\partial F_p}{\partial N_p} = K_B T \, ln\left(\frac{N_p \lambda_p^3}{V}\right) \quad (14)$$

**C- Free energy for neutral hydrogen**

Now, we turn our attention to the free energy of neutral hydrogen atoms which have both translational as well as internal free energies. From first principles one can write for the total (translational and internal) free energies of neutral hydrogen atoms the expression (see for example Ref. [28])

$$F_H = -N_H K_B T \left( ln\left(\frac{Q_{tot,H}}{N_H}\right) + 1 \right)$$
$$= N_H K_B T \left( ln\left(\frac{N_H \lambda_H^3}{V Q_{int,H}}\right) - 1 \right) \quad (15)$$

In the above equation, use has been made of the factorizability of the total partition function of the hydrogen atom, $Q_{tot,H}$, as in Potekhin's work, where $Q_{tot,H}$ is given by the product of the translational partition function, $Q_{trans,H}$, and the internal partition function, $Q_{int,H}$, that is;

$$Q_{tot,H} = Q_{trans,H} \, Q_{int,H} = \lambda_H^{-3} V \, Q_{int,H} \quad (16)$$

where $\lambda_H = h/\sqrt{2\pi m_H K_B T}$ is the thermal de Broglie wavelength of neutral hydrogen atoms. So far no assumption was made about $Q_{int,H}$. For the case of ideal plasma, the Maxwellian distribution can be used where

$$N_k = N_H \, g_k e^{\chi_k / K_B T} / Q_{int,H} \quad (17)$$

and the partition function or the normalization constant $Q_{int,H}$ is given by

$$Q_{int,H} = \sum_k g_k \, e^{\chi_k / K_B T} \quad (18)$$

with $\chi_k$ representing the unperturbed binding energy of the *k*th quantum state.

Since $N_H = \sum_k N_k = const.,$ and the quantity $N_H/Q_{int,H}$ is the same for all excited species *k*, then substituting from (17) into (15) gives

$$F_{id,H} = K_B T \sum_k N_k \left( ln\left(\frac{N_k \lambda_H^3}{V g_k e^{\chi_k/K_B T}}\right) - 1 \right) \quad (19)$$

However, if one considers the case of nonideal plasma, the Boltzmann distribution may be replaced by any real distribution of the form

$$N_k = N_H g_k \omega_k e^{\chi_k/K_B T} / Q_{int,H,\omega} \quad (17')$$

where $Q_{int,H,\omega}$ is the corresponding generalized IPF or normalization constant, defined as given by Potekhin in Ref. [5];

$$Q_{int,H,\omega} = \sum_k g_k \omega_k e^{\chi_k/K_B T} \quad (18')$$

Substituting from Eq. (17') and Eq. (18') into Eq. (15) one gets for the free energy of neutral hydrogen atoms in a nonideal plasma environment the expression

$$F_H^{non} = K_B T \sum_k N_k \left( ln\left(\frac{N_k \lambda_H^3}{V g_k \omega_k e^{\chi_k/K_B T}}\right) - 1 \right) \quad (19')$$

Recalling that the translational free energy for hydrogen atoms is given by

$$F_{trans,H} = K_B T \sum_k N_k \left( ln\left(\frac{N_H \lambda_H^3}{V}\right) - 1 \right) \quad (20)$$

one can subtract the translational free energy for neutral hydrogen atoms (Eq. (20)) from the above expressions for the total free energy of the neutral hydrogen for the case of ideal plasma (Eq. (19)) and for the case of nonideal plasma (Eq. (19')) to get for the internal free energy for both cases, in order, the following expressions

$$F_{int,H}^{id} = K_B T \sum_k N_k \, ln\left(\frac{N_k}{N_H \, g_k \, e^{\chi_k/K_B T}}\right) \tag{21}$$

and

$$F_{int,H}^{non} = K_B T \sum_k N_k \, ln\left(\frac{N_k}{N_H \, g_k \, \omega_k \, e^{\chi_k/K_B T}}\right) \tag{21'}$$

Eq. (21) is the same as Eq. (7) in Potekhin's paper, however, Eq. (21') is different as it has the factor $\omega_k$ in the denominator of the argument of the logarithm. Now, substituting from Eq. (17') into Eq. (21') onr gets

$$F_{int,H}^{non} = -K_B T \sum_k N_k \, ln(Q_{int,H,\omega}) \tag{22}$$

which differs from Eq. (15) in Potekhin's paper where Potekhin's expression has an extra term, $\sum_k N_k \, ln\,\omega_k$, in its right hand side. This extra term which has been interpreted as a contribution to the ideal-gas part of the entropy due to the correction $\omega_k$ to the probability that the *k*th state is occupied is clearly *incorrect* and should not be present in Eq. (15) in Potekhin's paper. A possible source for the erroneous appearance of this term in Eq. (15) in Potekhin's paper could be the substitution from Eq. (17') into Eq.(21) instead of Eq. (21') which represents an inconsistency in Potekhin's formulation.

Based on the assumed separability of the configurational free energy component, one can find the partial derivative of the total free energy of neutral hydrogen atoms with respect to $N_k$ as equal to the sum ($\partial F_{trans,H}/\partial N_k + \partial F_{int,H}^{id}/\partial N_k + \partial F_{conf}/\partial N_k$). Equivalently using the real distribution (Eq. (17') and Eq. (18')) instead of the Maxwellian distribution, the partial derivative of the total free energy of neutral hydrogen atoms with respect to $N_k$ can be expressed as the sum $(\partial F_{trans,H}/\partial N_k + \partial F_{int,H}^{non}/\partial N_k)$. Using the expressions for $F_{trans,H}$, $F_{int,H}^{id}$ and $F_{int,H}^{non}$ from Eq. (20), Eq. (21) and Eq. (21') one gets

$$\frac{\partial F_{trans,H}}{\partial N_k} = K_B T \, ln\left(\frac{N_H \lambda_H^3}{V}\right) \tag{23}$$

$$\frac{\partial F_{int,H}^{id}}{\partial N_k} = -K_B T \, ln(Q_{int,H}) \tag{24}$$

and

$$\frac{\partial F_{int,H}^{non}}{\partial N_k} = -K_B T \, ln(Q_{int,H,\omega}) - K_B T \sum_\alpha N_\alpha \frac{\partial}{\partial N_k} ln Q_{int,H,\omega} \tag{24'}$$

The minimization of the free energy, therefore, requires

$$\frac{\partial F}{\partial N_k} = \frac{\partial F}{\partial N_p} + \frac{\partial F}{\partial N_e} \tag{25}$$

At this stage, if one considers the two assumed equivalent cases; the case with the ideal free energies plus a separable configurational component and the equivalent case in which the nonideal free energies (including $Q_{int,H,\omega}$) is used, Eq. (25) for the first case can be written as

$$\left(\frac{\partial F_{conf}}{\partial N_k} - \frac{\partial F_{conf}}{\partial N_p} - \frac{\partial F_{conf}}{\partial N_e}\right) = -\frac{\partial F_{trans,H}}{\partial N_k} - \frac{\partial F_{int,H}^{id}}{\partial N_k} + \frac{\partial F_p}{\partial N_p} + \frac{\partial F_e^{clc}}{\partial N_e} + \frac{\partial \Delta F_e^{dgc}}{\partial N_e} \tag{26}$$

Upon substitution for the first four terms in the r.h.s. of Eq. (26) from Eq. (23), Eq. (24), Eq. (14) and Eq. (12), respectively, and using number densities in some places one arrives at

$$\sum_k n_k = \frac{n_p n_e}{2}\left(\frac{\lambda_p \lambda_e}{\lambda_H}\right)^3 \sum_k g_k e^{\chi_k/K_B T}$$
$$\times exp\left[-\frac{1}{K_B T}\left(\left[\partial/\partial N_k - \partial/\partial N_p - \partial/\partial N_e\right]F_{conf} - \partial\Delta F_e^{dgc}/\partial N_e\right)\right] \quad (27)$$

Equating the *k*th term of both sides of (27) gives

$$n_k = n_p n_e \left(\frac{\lambda_p \lambda_e}{\lambda_H}\right)^3 (g_k/2)\, exp\left[\frac{1}{K_B T}\left(\chi_k - \left[\partial/\partial N_k - \partial/\partial N_p - \partial/\partial N_e\right]F_{conf} - \partial\Delta F_e^{dgc}/\partial N_e\right)\right] \quad (28)$$

Performing similar steps except for using the sum ($\partial F_{trans,H}/\partial N_k + \partial F_{int,H}^{non}/\partial N_k$) in favor of the partial derivative of the free energy of neutral hydrogen-atoms with respect to $N_k$, leads to

$$\left(\frac{\partial F_{int,H}^{non}}{\partial N_k} - \frac{\partial F_{int,H}^{non}}{\partial N_p} - \frac{\partial F_{int,H}^{non}}{\partial N_e}\right) = -\frac{\partial F_{trans,H}}{\partial N_k} + \frac{\partial F_p}{\partial N_p} + \frac{\partial F_e^{clc}}{\partial N_e} + \frac{\partial \Delta F_e^{dgc}}{\partial N_e} \quad (26')$$

Upon using the expressions from Eq. (23), Eq. (24'), Eq. (14) and Eq. (12) for the partial derivatives in both sides and using number densities at some places one arrives at

$$\sum_k n_k = \frac{n_p n_e}{2}\left(\frac{\lambda_p \lambda_e}{\lambda_H}\right)^3 \sum_k g_k \omega_k e^{\chi_k/K_B T}$$
$$\times exp\left[-\sum_\alpha N_\alpha \left(\frac{\partial}{\partial N_k} - \frac{\partial}{\partial N_p} - \frac{\partial}{\partial N_e}\right) ln Q_{int,H,\omega} - \frac{1}{K_B T}\partial\Delta F_e^{dgc}/\partial N_e\right] \quad (27')$$

Equating the *k*th term of both sides of (27') gives

$$n_k = n_p n_e \left(\frac{\lambda_p \lambda_e}{\lambda_H}\right)^3 (g_k \omega_k/2) \times exp\left[+\frac{1}{K_B T}\left(\chi_k + \Delta\chi_k^{Q_\omega} + \Delta\chi_k^{dgc}\right)\right] \quad (28')$$

Where

$$\Delta\chi_k^{Q_\omega} = -N_H K_B T\left(\frac{\partial}{\partial N_k} - \frac{\partial}{\partial N_p} - \frac{\partial}{\partial N_e}\right) ln Q_{int,H,\omega}$$
$$= -K_B T \sum_k N_k \left(\frac{\partial}{\partial N_k} - \frac{\partial}{\partial N_p} - \frac{\partial}{\partial N_e}\right) ln \omega_k \quad (29)$$

and

$$\Delta\chi_k^{dgc} = -\frac{\partial \Delta F_e^{dgc}}{\partial N_e} \qquad (30)$$

The compact notations $\Delta\chi_k^{Q_\omega}$ and $\Delta\chi_k^{dgc}$ have been used here as their corresponding terms appear in Eq. (28') as if they were modifications or corrections added to $\chi_k$, however we deal with them here mathematically without any prejudiced physical interpretation at the moment, in an effort to avoid controversial argumentations raised in the literature (see for example Refs. [2,5]).

Comparing Eq. (28) and Eq. (28') one gets

$$\omega_k = exp\left[-\frac{1}{K_B T}\left[\partial/\partial N_k - \partial/\partial N_p - \partial/\partial N_e\right]F_{conf} + N_H\left[\partial/\partial N_k - \partial/\partial N_p - \partial/\partial N_e\right]\ln Q_{int,H,\omega}\right] \qquad (31)$$

Eq. (31) indicates that Potekhin's expression for $\omega_k$ (Eq. (5)) cannot be obtained unless

$$\left(\frac{\partial}{\partial N_k} - \frac{\partial}{\partial N_p} - \frac{\partial}{\partial N_e}\right)\ln Q_{int,H,\omega} = \frac{1}{Q_{int,H,\omega}}\sum_\alpha g_\alpha e^{\chi_\alpha/K_B T}\left(\frac{\partial}{\partial N_k} - \frac{\partial}{\partial N_p} - \frac{\partial}{\partial N_e}\right)\omega_\alpha = 0 \qquad (32)$$

which represents a restriction that has never been declared in Potekhin's paper and the formulation (5) was confusingly introduced as a general result which is not correct. Taking Eq. (5) into consideration, it becomes clear that the above restriction is equivalent to requiring the interaction term or the configurational component of the free energy to be linear in the occupation numbers which is the same restriction imposed on the Hummer-Mihalas formulation. That is equivalent to having $\omega_\alpha$ independent of the occupation numbers of the particle species which, in turn, contradicts the required dependence of $\omega_\alpha$ on the density or occupation numbers. The partition function calculated this way will be density-independent, lacks a scheme to terminate the sum over states and will be subject to the same criticism raised against the density-independent partition functions. Yet, a model for $F_{conf}$ that satisfies this restriction cannot be easily accepted or justified, particularly with the inclusion of the long range Coulomb forces among charged particles.

**IV– A CONSISTENT FORMULATION AND THE SOLUTION OF THE INVERSE PROBLEM**

Equation (31) represents an identity that is valid for occupational probability formalism of the form (2) and any associated proposed form of a separable configurational component of the free energy (to be added to the ideal free energy). This identity can be simplified to the form

$$\left[\frac{\partial}{\partial N_k} - \frac{\partial}{\partial N_p} - \frac{\partial}{\partial N_e}\right] F_{conf} = -K_B T \ln \omega_k + \Delta \chi_k^{Q_\omega} = \Delta \chi_k^Q \tag{33}$$

Before going deeper into the analysis, it is helpful to recognize that if $\omega_k$ is known (henceforth we refer to this case as the inverse problem) one can easily calculate the occupation numbers by solving the set of Eqs. (28') supplemented by the stoichiometric constraints. The remaining part, however, is to find the corresponding modifications to the thermodynamic properties to complete the self-consistent formulation of the problem. On the other hand, if $F_{conf}$ is known (henceforth we refer to this case as the forward problem), obtaining the modifications to the thermodynamic properties can be straightforwardly performed through evaluating the partial derivatives of $F_{conf}$. However, obtaining $\omega_k$ consistently in this case and consequently the calculation of the occupation numbers does not seem to be a manageable task except for the trivial case of density-independent $\omega_k$ or equivalently the case of linear dependence of $F_{conf}$ on the occupation numbers.

Having such difficulties associated with the general solution of the forward problem, the formulation of a complete solution of the inverse problem becomes exceptionally valuable and inevitable. It has been stated above that once a scheme to establish finite IPFs is developed in advance; it can be incorporated into and used to solve the set of Eqs. (28') supplemented by the stoichiometric constraints to obtain the occupation numbers of all species. In principle, one can add to the total free energy a separable interaction term, $F_C$ (Coulombic excess free energy) that can not be exclusively responsible for the termination of the partition function with *non-perturbed energy levels*. This separable term, *if any*, would lead to correction terms to the pressure $\Delta P_{F_C}$ and the internal energy $\Delta U_{F_C}$. Fortunately, the remaining part of obtaining the corresponding corrections to other thermodynamic properties can be achieved straightforwardly through evaluating the partial derivatives of the expressions of the free energy components with the nonideal internal component where the pressure and the internal energy are given respectively by

$$P = -\left(\frac{\partial F}{\partial V}\right)_{T,\{N_k\}} = \left(2K_BT/\Lambda_e^3\right)I_{3/2}(\mu_e/K_BT) + \frac{(N_H+N_p)K_BT}{V} + \Delta P_{F_C} + \Delta P_{Q_\omega} \quad (34)$$

and

$$\begin{aligned}U &= -T^2\left(\frac{\partial F/T}{\partial T}\right)_{V,\{N_k\}} \\ &= \left(3K_BTV/\Lambda_e^3\right)I_{3/2}(\mu_e/K_BT) + \frac{3(N_H+N_p)K_BT}{2V} - \sum_k N_k\chi_k + \Delta U_{F_C} + \Delta U_{Q_\omega}\end{aligned} \quad (35)$$

where $\Delta P_{F_C} = -\partial F_C/\partial V$, $\Delta P_{Q_\omega} = K_BT\sum_k N_k\frac{\partial}{\partial V}\ln\omega_k$ and $\Delta U_{F_C} = -T^2\partial(F_C/T)/\partial V$ and

$\Delta U_{Q_\omega} = K_BT^2\sum_k N_k\frac{\partial}{\partial T}\ln\omega_k$. It has to be noted that following Potekhin's notations, the

reference level for the binding energies used above is that of a free electron or the case in which the hydrogen atom is ionized. Accordingly, the third term in Eq. (35) represents the recombination energy emitted from the system to form the bound states *k*. The last term in each of the equations (34) and (35) is the consequence of the scheme used to terminate the partition function.

The set of Eqs (28', 34 and 35) along with the constraints of electroneutrality and conservation of nuclei together with the scheme used to establish finite bound-state partition functions represent the complete set of equations to be used to consistently calculate the occupation numbers and thermodynamic properties of the nonideal plasma system under consideration.

Although the above methodology fully solves the inverse problem for any form of $\omega_k$ one may ask for curiosity; Given the form of $\omega_k$, is it always possible to represent the occupational probability, $w=g\omega$, by a corresponding *separable* configurational term to be added to the ideal free energy?

In the above sections we have seen that this is possible for the case of occupation probability that is independent of the occupation numbers with the associated $F_{conf}$ linear in the occupation numbers. That is apparently a trivial case (as far as the truncation of the IPF is concerned) since it does not *exclusively* satisfy the essential drive of having a density-dependent finite IPF.

Before one tries to answer the above question, it may be illuminating to bring into attention that it is not necessary to have a separable configurational term in order to have the scheme used to establish finite IPF self-consistently reflected in the modifications of the free energy. As a matter of fact the internal structure of a composite particle is generally affected by the surrounding and therefore the separation of the configurational component from the internal component of the free energy may not be always possible. In the appendix we analytically answer this question, showing the separability criterion and the methodology for deriving $F_{conf}$ in this case.

**V. ILLUSTRATIVE EXAMPLE**

The main thrust of section *IV* was the introduction of a consistent mathematical formulation of the problem of the calculation of thermodynamic properties and the establishment of the finite bound state partition function within the chemical picture. As has been shown above, the consistent establishment of finite IPFs in the inverse scheme necessitates the introduction of new correction terms to the ionization potentials, $\Delta\chi_k^{Q_\omega}$, plasma pressure, $\Delta P_{Q_\omega}$, and internal energy, $\Delta U_{Q_\omega}$, as consequences of the scheme used, in advance, to terminate the IPF. In this section, an illustrative example is worked out to show both simplicity and effectiveness of the proposed method in terms of the significance of these corrections on top of being essential for consistency. In the present example we use the model for the occupational probability given by Salzmann in Ref. [29] where the occupational probability is given by

$$\omega_k = exp\left[-\left(\frac{\Delta\chi_H}{\chi_{H,k}}\right)^3\right] \qquad (38)$$

where $\chi_{H,k}$ is the energy of level *k* measured from the continuum and $\Delta\chi_H$ is the lowering of the ionization potential of the neutral hydrogen. The above expression shows clearly that $\omega_k$ goes to unity as $\Delta\chi_H$ goes to zero which is the case of ideal plasma and that $\omega_k$ smoothly goes to zero as $\Delta\chi_H$ exceeds (- $\chi_{H,k}$) as desired for the termination of the IPF. The lowering of the ionization energies has been expressed in the form

$$\Delta\chi_\zeta = -C\frac{(\zeta+1)e^2}{4\pi\varepsilon_0 R_0} \qquad (37)$$

where C is a constant, $R_0 = (3/4\pi n)^{1/3}$ and $n=n_H+n_p$ is the number density of heavy particles. The corresponding Coulomb correction to the free energy (separable part of the configurational free energy) can be derived as shown in Ref. [30] to be

$$F_C = -\sum_{\zeta}^{\zeta_{max}} n_\zeta \frac{C}{2} \frac{\zeta^2 e^2}{4\pi\varepsilon_0 R_0} \tag{38}$$

In a more rigorous theory, Eqs. (36)-(38) may be derived based on the same physical model of particle interactions.

Using the above expressions in (36)-(38) one can find all corrections for the pressure and internal energies in Eqs. (34) and (35). It is not difficult to show that $\Delta\chi_k^{Q_\omega}$ and $\Delta U_{Q,\omega}$ are both zeros for this particular case. The only consequence of using the occupational probability in Eq. (36) is therefore the correction to the pressure $\Delta P_{Q_\omega}$ or the pressure of shell compression.

Figure 1 shows the individual components of the pressure due to free electrons $P_e$, Coulombic correction $\Delta P_{F_C}$, shell compression $\Delta P_{Q_\omega}$ and the total pressure of a 20000 K hydrogen plasma all normalized to $nK_BT$. The component of the pressure due heavy particles (not shown in the figure) is understood to be unity in this case. As expected, the Coulombic correction term (separable part of thr configurational free energy) contributes a negative component to the pressure which becomes significant at high densities. The component of the free electrons' pressure increases significantly at high densities, as the degree of degeneracy increases, preventing the collapse of the plasma in this case. It is a common knowledge that a degenerate electron gas has a *much higher* pressure than that which would be predicted by classical physics. This is an entirely quantum mechanical effect due to the fact that identical fermions cannot get significantly closer together than a de Broglie wave-length without violating the Pauli exclusion principle. As it can be seen from the figure, the shell compression component $\Delta P_{Q_\omega}$ has a non-negligible effect and contributing a significant positive component of the computed total pressure. This component is however essential for the self-consistency of the set of thermodynamic functions along with the calculated occupation numbers. Figure 2 shows the total pressure for the same plasma system with and without the shell compression component $\Delta P_{Q_\omega}$ confirming the quantitative importance of this component as discussed above. For completeness we provide in Figures 3 and 4, respectively a set of isotherms of the pressure and degree of ionization of hydrogen plasma as functions of the heavy particle number density. The effect of using Potekhin's approximation for electron degeneracy (Eq. 8) on the calculation of a 20000 K hydrogen plasma is shown in Fig. (5). Apparently the effect becomes significant at high densities where degeneracy becomes very important.

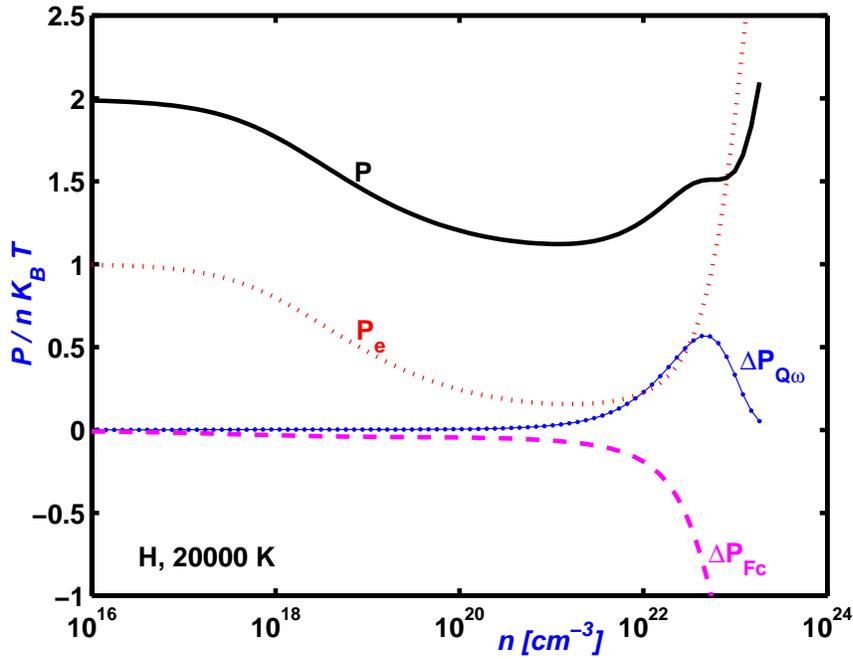

**Figure 1** Normalized components of free electrons' pressure, $P_e$, Coulombic correction term, $\Delta P_{F_C}$, and shell compression term, $\Delta P_{Q_\omega}$, along with the total pressure of a 20000 K hydrogen plasma as functions of the number density of heavy particles.

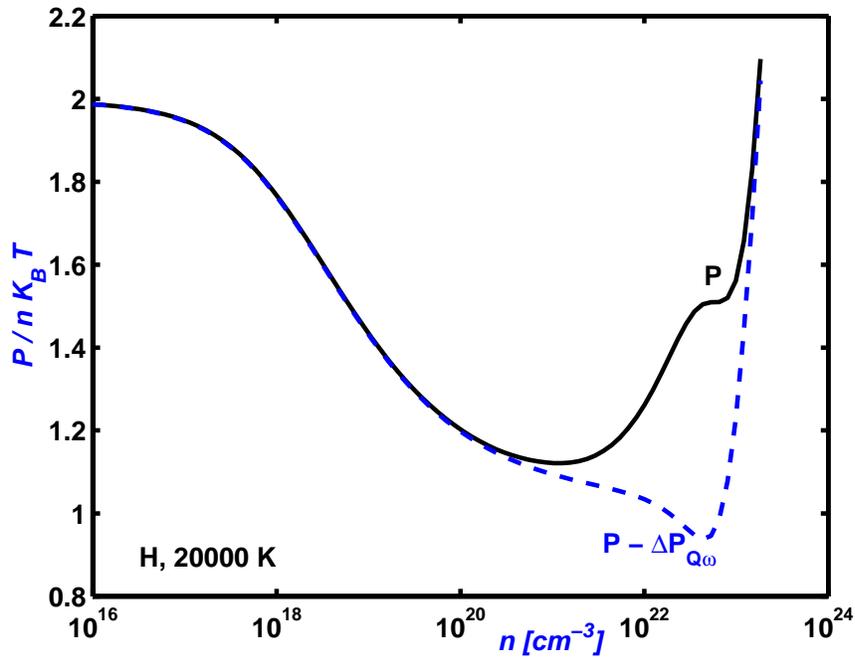

**Figure 2** Normalized total pressure of a 20000 K hydrogen plasma with and without the shell compression component, $\Delta P_{Q_\omega}$.

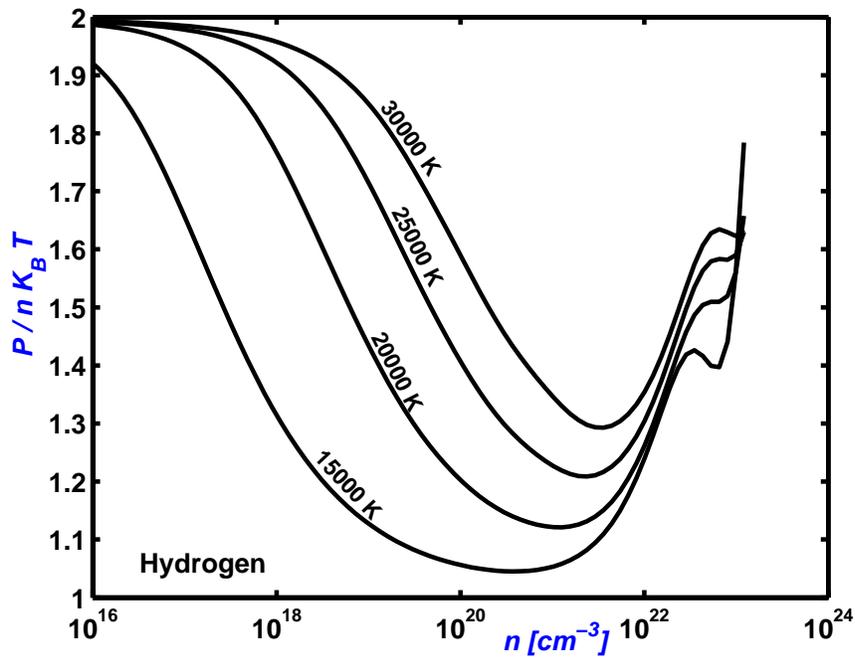

**Figure 3** Isotherms of normalized total pressure of hydrogen plasma as functions of the heavy particle number density.

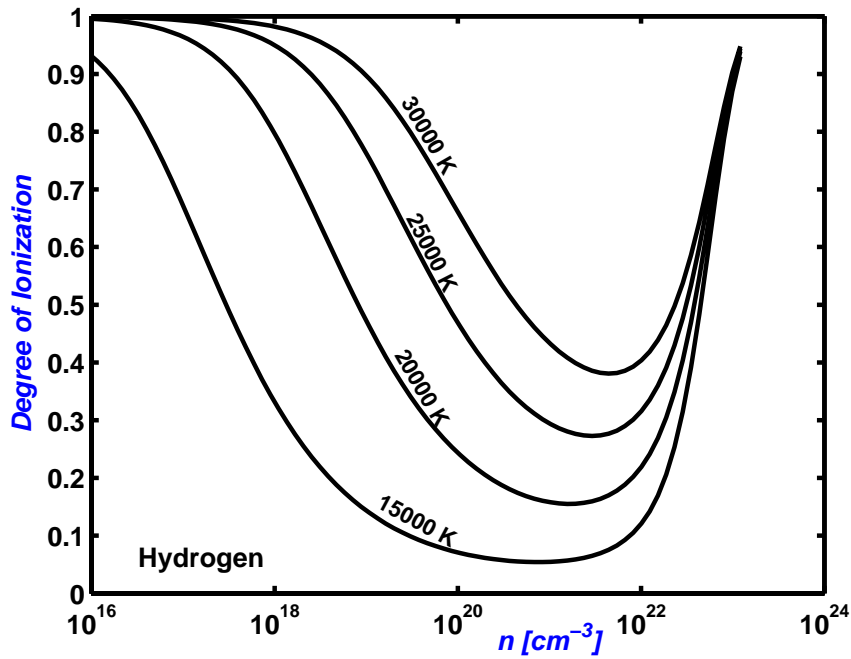

**Figure 4** Isotherms of the degree of ionization of hydrogen plasma as functions of the heavy particle number density

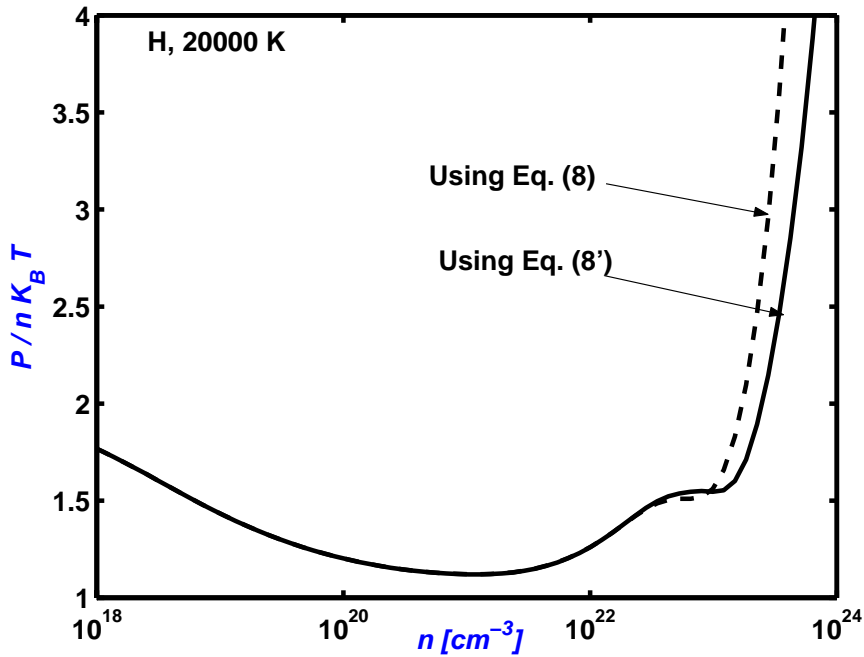

**Figure 5** Effect of using Potekhin's approximation for electron degeneracy (Eq. 8) on the calculation of the pressure of a 20000 K hydrogen plasma.

## VI. SUMMARY AND CONCLUSIONS

The problem of consistent calculation of equilibrium thermodynamic properties and the establishment of finite bound-state partition functions in nonideal hydrogen plasma systems is investigated within the chemical picture. The present exploration clears ambiguities and gives a better understanding of the problem on top of pointing out inaccuracies and inconsistencies buried in widely used models in the literature. A fully consistent formulation of the problem in terms of the solution of the inverse problem is introduced. The solution can be summarized as follows; a) based on physical basics, a scheme for establishing finite bound-state partition functions can be developed or adopted in advance, b) the developed or adopted model for the establishing of finite bound-state partition functions can be incorporated into the set of Eqs. (28'), c) the set of Saha-like equations (28') supplemented by the constraints of electroneutrality and conservation of nuclei can therefore be solved using available algorithms capable of solving the problem with relative simplicity and high accuracy, and d) the fundamental thermodynamic properties ($P$ and $U$) can be calculated using Eq. (36) and Eq. (37) while other thermodynamic properties, if needed, can be deduced from ($P$ and $U$) using thermodynamic relations and identies. The calculated equation-of-state and thermodynamic properties are fully consistent with the population numbers calculated from the solution of Eqs. (28') ensuring a consistent formulation of the problem. An illustrative example is worked out showing simplicity and effectiveness of the proposed consistent formulation and the importance of terms, essential for thermodynamic consistency, which are commonly neglected by other authors in the literature.

## Acknowledgments

The author would like to thank Prof. T. Thiemann from the Chemistry Dept., UAE University for the time and efforts he courteously devoted in translating Fermi's original article (reference 20) from German to English.

This work is supported by the UAE University, contract 05-02-2-11/08.

## Appendix

**A criterion of the separability of the configurational component of the free energy function**

Starting from Eq. (33), one can write

$$\Delta \chi_k^Q = \left[ \frac{\partial}{\partial N_k} - \frac{\partial}{\partial N_p} - \frac{\partial}{\partial N_e} \right] F_{conf} \tag{A-1}$$

The assumed separability of the configurational component of the free energy enables one to calculate its differential which can be expressed as

$$d F_{conf} = \sum_k \frac{\partial F_{conf}}{\partial N_k} dN_k + \frac{\partial F_{conf}}{\partial N_p} dN_p + \frac{\partial F_{conf}}{\partial N_e} dN_e \tag{A-2}$$

However, considering the constraint of electroneutrality $N_p = N_e$ Eqs (A-1) and (A-2) can be written as

$$\frac{\partial F_{conf}}{\partial N_k} = \Delta \chi_k^Q + 2 \frac{\partial F_{conf}}{\partial N_p} \tag{A-1'}$$

And

$$\begin{aligned} d F_{conf} &= \sum_k \left( \Delta \chi_k^Q + 2 \frac{\partial F_{conf}}{\partial N_p} \right) dN_k + 2 \frac{\partial F_{conf}}{\partial N_p} dN_p \\ &= \sum_k \Delta \chi_k^Q \, dN_k + 2 \frac{\partial F_{conf}}{\partial N_p} \sum_k \left( dN_k + dN_p \right) \\ &= \sum_k \Delta \chi_k^Q \, dN_k \end{aligned} \tag{A-2'}$$

Note that we made use of the fact that $(dN_k + dN_p) = dN_{tot} = 0$. The final result of Eq. (A-2') indicates that $F_{conf}$ could be considered effectively and exclusively as a function of $\{N_k\}$ only with

$$\frac{\partial F_{conf}}{\partial N_k} = \Delta \chi_k^Q \tag{A-3}$$

In general, the configurational component of the free energy is a function of $V$, $T$, $N_e$ and the populations of protons and different exited states of neutral hydrogen atoms. However since $V$ and $T$ are fixed, $N_e$ and $N_p$ are fully determined from the excited states' occupation numbers (occupation numbers of neutral species) through electroneutrality and conservation of nuclei, it becomes clear that the excited species are the only independent variables in $F_{conf}$ which is in complete agreement with the result of Eq. (A-3).

Bearing in mind that the separable $F_{conf}$ must be a single-valued or point function of the equilibrium state of the assembly, $dF_{conf}$ must be an *exact* differential which requires the scheme used to establish finite internal partition functions to satisfy the following *self-consistency* or *integrability* criterion;

$$\frac{\partial}{\partial N_\alpha} \Delta \chi_k^Q = \frac{\partial}{\partial N_k} \Delta \chi_\alpha^Q \tag{A-4}$$

If the condition (A-4) is satisfied, one can integrate the differential in Eq. (A-2') to obtain the configurational component of the free energy $F_{conf}$.

Having verified that $dF_{conf}$ is an exact differential one can proceed and integrate it choosing any path between the initial values of the variables and their final values. One possible path can be constructed through increasing the variables (excited states occupation numbers) in the same ratio (see for example [30]). Denoting the fraction of the final densities, which the excited particles have at any stage of integration, by λ then we have

$$F_{conf} = N_{tot} \sum_{k} \alpha_k \int_0^1 \Delta \chi_k^Q(\lambda) \, d\lambda \tag{A-5}$$

where $N_{tot}=N_H+N_p$ is the total number of heavy particles, $\alpha_k=N_k/N_{tot}$ is the molar fraction of the $k$-excited species and $F_{conf}(\lambda=0)=0$.

Another simple alternative path can be taken through sequentially increasing the variables (numbers of excited species) from their initial values (corresponding to $F_{conf}=0$) to their final values. This later path can be mathematically described and articulated as

$$F_{conf} = N_{tot} \sum_{k} \int_{\alpha_{i<k},\alpha_{i\geq k}=0}^{\alpha_{i\leq k},\alpha_{i>k}=0} \Delta \chi_k^Q(\alpha_{i\leq k}, \alpha_{i>k}=0) \, d\alpha_k \tag{A-6}$$

where $\alpha_k = N_k/(N_H + N_p)$ and $\Delta \chi_k^Q$ is given by.

$$\begin{aligned}
\Delta \chi_k^Q &= -K_B T \ln \omega_k - N_H K_B T \left( \frac{\partial}{\partial N_k} - \frac{\partial}{\partial N_p} - \frac{\partial}{\partial N_e} \right) \ln Q_{int,H,\omega} \\
&= -K_B T \ln \omega_k - \frac{N_H K_B T}{Q_{int,H,\omega}} \left( \frac{\partial}{\partial N_k} - \frac{\partial}{\partial N_p} - \frac{\partial}{\partial N_e} \right) Q_{int,H,\omega}
\end{aligned} \tag{A-7}$$

Although the integration in Eq. (A-5) or Eq. (A-6) can be evaluated analytically for some cases giving rise to exact closed form expressions for the configurational component of the free energy, in several other cases numerical evaluation may become unavoidable. In all cases, the corrections to thermodynamic functions can be derived from the calculated $F_{conf}$ using standard thermodynamic relations, which generally depend on the derivatives of $F_{conf}$ with respect to the independent state variables, namely volume, $V$, and temperature, $T$.

**REFERENCES**

1. G. M. Harris, J. E. Roberts, and J. G. Trulio, Phys. Review, 119 (6), 1832 (1960)

2. D. G. Hummer and Dimitri Mihalas, APJ 331:794-814 (1988)

3. D. Mihalas, W. Däppen, and D. G. Hummer, APJ 331, 815-825 (1988)

4. A. Y. Potekhin and G. Chabrier, Y. A. Shibanov, PRE, 60, 2193-2208 (1999)

5. A. Y. Potekhin, PoP 3 (11), 4156-4165 (1996)

6. Mofreh R. Zaghloul, PRE, 69, 026702 (2004).

7. Mofreh R. Zaghloul, IEEE Trans. *Plas. Sci.*, Vol. 33, No. 6 1973-1983 (2005).

8. Carl A. Rouse, APJ 272, 377-379 (1983).

9. Ebeling W., Kraeft W. D., Kremp D. and Röpke G., APJ 290:24-27 (1985)

10. Rogers F. J., APJ 310:723-728 (1986)

11. Däppen, W., Anderson, L. S., & Mihalas, D., APJ 319, 195 (1987)

12. Däppen, W., Nayafonov, A., APJ Supp. Series, 127, 287-292 (2000)

13. Hans-Werner Drawin and Paul Felenbok, Data for plasmas in local thermodynamic equilibrium, Gautheir-Villars Paris (1965)

14. H. N. Olsen, Physcal Review, 124, 6, 1703-1708 (1961)

15. M. Capitelli and G. Ferraro, SAB: Atomic Spectroscopy 31 (5) 323-326 (1976)

16. O. Cardona, E. Simonneau, and L. Crivellari, Revista Mexicana De Fisca 51 (5), 476-481 (2005)

17. D. Bruno, M. Capitelli, C. Catalfamo, and A. Laricchiuta, 28$^{th}$ ICPIG, July 15-20 Prague, Czeck Republic, 235-237 (2007)

18. Yu. V. Arkhipov, F. B. Baimbetov, and A. E. Davletov, CPP 47, 4-5, 248-252 (2007)

19. Gianpiero Colonna and Mario Capitelli, SAB: Atomic Spectroscopy 64, 863-873 (2009)

20. E. Fermi, Z. Phys., 26, 54 (1924)

21. V. E. Zavlin, G. G. Pavlov and Yu. A. Shibanov, Astron. & Astro. Trans., 4 (4), 307 - 312 (1994)

22. T. Nishikawa, APJ 532, 670-672 (2000)

23. Kilcrease, David Hakel, Peter Abdallah, Joseph Magee, Norman Mazevet, Stephane and Sherrill, Manolo, "An Occupation-Probability-Formalism Equation-of-State For New Opacity Calculations," American Physical Society, 45th Annual Meeting of the Division of Plasma Physics, October 27-31, 2003, Albuquerque, New Mexico, MEETING ID: DPP03, abstract #QP1.167, (APS Meeting Abstracts, 1167P)

24. Hakel, Peter and Kilcrease, David P, AIP Conference Proceedings, 730 (1), 14th APS topical conference on atomic processes in plasmas, Santa Fe, NM (United States), 19-22 Apr (2004)

25. Alexander Y. Potekhin, PRE 72, 046402 (2005)

26. Mofreh R. Zaghloul, PRE 79 (3), 037401 (2009).

27. W. Ebeling, A. Förster, V. Fortov, V. Gryaznov, A. Polishchuk, Thermophysical Properties of Hot Dense Plasmas, (Teubner Verlagsgesellschaft, Stuttgart 1991)

28. R. E. Sonntag and G. Van Wylen, Introduction to Thermodynamics: Classical and Statistical, second edition (John Wiley & Sons Inc., USA, 1982)

29. David Salzmann, Atomic Physics in Hot Plasmas (Oxford University Press, New York, Oxford 1998)

30. Mofreh R. Zaghloul, APJ 699, 885-891 (2009).